\documentstyle[12pt,epsf,epsfig,psfig]{article}

\def\J{$J/\psi$}
\def\j{J/\psi}
\def\i{\psi}

\def\P{$\psi'$}
\def\p{\psi'}

\def\C{c{\bar c}}

\def\q{q{\bar q}}

\def\_{\!-\!}

\def\be{\begin{equation}}
\def\ee{\end{equation}}

\def\lsim{\raise0.3ex\hbox{$<$\kern-0.75em\raise-1.1ex\hbox{$\sim$}}}
\def\gsim{\raise0.3ex\hbox{$>$\kern-0.75em\raise-1.1ex\hbox{$\sim$}}}

\def\NP{{ Nucl.\ Phys.\ }}
\def\PL{{ Phys.\ Lett.\ }}
\def\PR{{ Phys.\ Rev.\ }}

\def\PRL{{ Phys.\ Rev.\ Lett.\ }}

\begin{document}


\vskip 1.5cm

\centerline{\large{\bf Photoproduction Constraints on \J-Nucleon
Interactions}}

\vskip 1.0cm

\centerline{\bf K.\ Redlich$^{1,2,3}$, H.\ Satz$^1$ and G.\ M.\
Zinovjev$^{1,4}$}

\vskip 1.0cm
\noindent
1: Fakult\"at f\"ur Physik, Universit\"at Bielefeld, D-33501 Bielefeld,
Germany \par\noindent
2:  Gesellschaft f\"{u}r  Schwerionenforschung, D-64220  Darmstadt, Germany \par\noindent
3: Institute of Theoretical Physics, University of Wroc\l aw,
PL-50204 Wroc\l aw, Poland \par\noindent
4: Bogolyubov Institute of Theoretical Physics, Academy of Sciences,
\par\noindent ~~~~UA-252143 Kiev, Ukraine

\vskip 0.6cm

\centerline{\bf Abstract:}

\bigskip

Using \J~and open charm photoproduction data, we apply the vector
meson dominance model to obtain constraints on the energy dependence
of the inelastic \J-nucleon cross section. Predictions of short
distance QCD are in accord with these constraints, while recently
proposed hadronic models for \J~dissociation strongly violate them.

\vskip 1cm

The energy dependence of the inelastic \J-nucleon cross section
$\sigma^{in}_{\i N}(s)$ is of great importance in understanding
\J~suppression as signature for colour deconfinement in high energy
nuclear collisions \cite{Matsui}. Calculations based on short distance
QCD predict a strong threshold damping of $\sigma^{in}_{\i N}(s)$,
due to the suppression of high momentum gluons by the gluon
distribution function in nucleons \cite{B-P}-\cite{KS3}; this damping
persists also when finite target mass corrections are taken into
account \cite{KSSZ1}. In contrast to such QCD studies, several
recently proposed models based on hadron exchange suggest large
threshold values of $\sigma^{in}_{\i N}(s)$ \cite{M&M} - \cite{Ko}.
The aim of this note is to show that available \J~and open charm
photoproduction data can do much to clarify the situation.

\medskip

The existing empirical information on \J-hadron interactions comes from
pho\-to\-pro\-duction and the vector meson dominance model (VMD)
\cite{Sakurai}, which relates $e^+e^- \to \i,~\gamma N \to \i N$ and
$\i\_N$ data \cite{HLW}. It is based on the assumption that
fluctuations of the photon into quark-antiquark pairs are dominated by
the corresponding hadronic resonances. As a result, the diffractive
\J-photoproduction cross section is related to elastic $\i\_N$
scattering,
\be
\sigma(\gamma N \to \i N) = \left( {4 \pi \alpha \over \gamma_{\i}^2}
\right) \sigma_{\rm el}^{\i N}. \label{1}
\ee
Here $\gamma_{\i}$ is determined by the \J-decay into $e^+e^-$,
\be
\Gamma(e^+e^- \to \i) = {\alpha^2 \over 3}\left( {4 \pi \over
\gamma_{\i}^2} \right) M_{\i},\label{2}
\ee
with $\Gamma(e^+e^- \to \i) = 5.26 \pm 0.37$ keV \cite{PD}.
Furthermore, the optical theorem leads to
\be
\left( {d\sigma (\gamma N \to \i N) \over dt} \right)_{t=0} =
{(1 + \rho^2) \over 16 \pi} \left( {4 \pi \alpha \over \gamma_{\i}^2}
\right) (\sigma_{\rm tot}^{\i N})^2 , \label{3}
\ee
where $\rho=[{\rm Re}~M(s)/{\rm Im}~M(s)]$ is the ratio of real to
imaginary part of the $\i\_N$ forward scattering amplitude. This vanishes at
high energy, so that then Eq.\ (\ref{3}) relates the total $\i\_N$
cross section to forward \J-photoproduction.

\medskip

The first experimental measurements of the \J -photoproduction cross
section had already shown it to be very small compared to the 
corresponding cross sections for conventional vector mesons $\rho, 
\omega$ and $\phi$ \cite{Knapp}. One of the first explanations of 
this result had invoked the smallness of the Pomeranchuk pole residue 
for the \J, i.e., the total cross section of $\i\_N$-interaction should 
be small, and the interaction of $\pi$ and \J~was argued to be quite weak
\cite{VIZ,STW}. Moreover, it was concluded there the \J~interaction
with hadrons should be dominated by charmed particle production.

\medskip

Today there exist quite good data.
For c.m.s.\ energy $\sqrt s \simeq 20$ GeV (corresponding to a photon
energy of about 200 GeV), the forward photoproduction cross section is about
100 nb/GeV$^2$ \cite{HLW}. Assuming that here $\rho \simeq 0$, and using
the quoted value for $\Gamma(e^+e^- \to \i)$, we get
$\sigma_{\rm tot}^{\i N} \simeq 1.7$ mb. Geometric arguments, which also assume
$\rho=0$, predict $\sigma_{\rm tot}^{\i N}/\sigma_{\rm tot}^{NN} \simeq
(r_{\i}/r_N)^2$ \cite{P-H}. With $r_{\i} \simeq 0.2$ fm, $r_N \simeq
0.85$ fm and $\sigma_{\rm tot}(NN) \simeq 40$ mb, this gives
$\sigma_{\rm tot}^{\i N} \simeq 2.2$ mb. Thus both VDM and geometric
considerations lead to a total high energy $\i\_N$ cross section
around 2 mb.

\medskip

At $\sqrt s \simeq 20$ GeV, $\sigma(\gamma N \to \i N) \simeq 17.5$ nb
\cite{HLW}; using Eq.\ (\ref{1}), we obtain
\be
\sigma_{\rm el}^{\i N}  \simeq 25~\mu{\rm b} \label{4}
\ee
for the elastic $\i\_N$ cross section at this energy. Hence the
high energy ratio of elastic to total $\i\_N$ cross sections is with
\be
{\sigma_{\rm el}^{\i N} \over \sigma_{\rm tot}^{\i N}} \simeq
{1 \over 70} \label{5}
\ee
very much smaller than that for the interaction of light hadrons; the
corresponding $\pi\_N$ ratio is an order of magnitude larger.
At high energy, the total $\i\_N$ cross section is thus strongly
dominated by inelastic channels; for the \J, it is appearently much
more difficult to survive high energy interactions than it is for
hadrons consisting of light quarks, so that most of
$\sigma_{\rm tot}^{\i N}$ consists of open charm production. 
This is in accord with the Okubo-Zweig-Iizuka (OZI) rules, which
forbid the \J~as $\C$ bound state to annihilate into ordinary light
hadrons and hence lead to charmed meson production. Such behaviour
is also a natural consequence of partonic interactions, rather 
than black disc absorption.

\medskip

Since Eq.\ (\ref{3}) determines the total cross section only modulo
$(1+\rho^2)^{1/2}$, additional information is needed to determine
$\sigma_{\rm in}^{\i N}(s)$. This is provided by the photoproduction of
open charm, which we denote by $\sigma(\gamma N \to \C)$; it
is empirically obtained by measuring $D$ and $D^*$ production. From
VMD, we expect
\be
\sigma(\gamma N \to \C) \simeq
\left( {4 \pi \alpha \over \gamma_{\i}^2} \right)
\sigma_{\rm in}^{\i N}. \label{6}
\ee
Before applying this relation, the role of other vector mesons must be
clarified. Intermediate light quark states, such as $\rho$ or $\omega$,
could also produce open charm. Data on the cross section for open
charm hadroproduction, in accord with perturbative calculations \cite{HPC},
give some 10 - 20 $\mu$b at $\sqrt s \simeq 20$ GeV.
This is to be compared to $\sigma_{\rm tot}^{\i N} \simeq 2$ mb at the
corresponding energy, keeping in mind the ratio of the photon
couplings $\gamma_{\rho}^{-2}/ \gamma_{\i}^{-2} \simeq 5.18$.
Light vector mesons therefore contribute to open charm photoproduction
at most on a 5 \% level.

\medskip

Further contributions could come from higher $\C$ resonances, such as
the \P. These are in fact also negligible, but for a different
reason. VMD implicitly assumes that the fluctuations of a real photon
into a $\q$ pair are comparable in size to the relevant vector mesons.
For light quarks and light mesons, this is the case, since both are
of typical hadronic scale. For $\gamma \to \C$, the scale is very
much smaller, but it is also correspondingly smaller for the \J, with
both around 0.1 - 0.2 fm; hence VDM still makes sense. The higher $\C$
vector mesons are much larger than the $\C$ fluctuation, however, and
so for them VMD `fails' \cite{STW,H-K}. This can be checked by
considering the ratio of `elastic' \J~to \P~photoproduction. From VMD
and the optical theorem, one expects
\be
{\sigma(\gamma N \to \p N) \over \sigma(\gamma N \to \i N)} =
\left({M_{\i} \over M_{\p}}\right)
\left({\Gamma(e^+e^- \to \p) \over \Gamma(e^+e^- \to \i)}\right)
\left( {\sigma_{\rm tot}^{\p N} \over \sigma_{\rm tot}^{\i N}} \right)^2.
\label{7}
\ee
Geometric arguments \cite{P-H} suggest $\sigma_{\rm tot}^{\p N}/
\sigma_{\rm tot}^{\i N} \simeq 4$, since the radius of the $2S$ state is
more than twice that of the $1S$. Inserting the corresponding masses and
decay widths, the ratio $\sigma(\gamma N \to \p N)/\sigma(\gamma N \to
\i N)$ is predicted to be 5.5. Photoproduction data \cite{H1}, in contrast,
give a ratio of $0.15 \pm 0.03$, more than a factor 30 smaller.
Evidently the \P~can therefore also be neglected as an intermediate
state in open charm photoproduction.\footnote{In $e^+e^-$ collisions,
the \P~continues to appear in VDM strength, so that its decoupling in
photoproduction can also be considered as an effect of the
exptrapolation from highly virtual to real photons \cite{STW}.}

\medskip

As a final consistency check, we can see if the $\sigma_{\rm in}^{\i N}$
determined by Eq.\ (\ref{6}) from open charm photoproduction
indeed converges at high energies to the $\sigma_{\rm tot}^{\i N}$
obtained from forward \J~photoproduction by Eq.\ (\ref{3}).
It will be found shortly that this is indeed the case.

\medskip

We thus use the data for open charm photoproduction \cite{Nash,Ali}
and Eq.\ (\ref{6}) to determine the energy dependence of
$\sigma_{\rm in}^{\i N}(s)$, while \J~photoproduction \cite{HLW}
and Eq.\ (\ref{1}) gives that of $\sigma_{\rm el}^{\i N}(s)$.
The results are shown in Fig.\ 1, together with the data for
$(1 + \rho^2)^{1/2} \sigma_{\rm tot}^{\i N}(s)$ as obtained
from forward \J~photoproduction through VMD and the optical theorem
(Eq.\ (\ref{3})). We note that at high energy, where we expect
$\rho \to 0$, $\sigma_{\rm in}^{\i N}(s)$ indeed approaches
$\sigma_{\rm tot}^{\i N}(s)$, so that the consistency check just
mentioned is satisfied. The curves shown in Fig.\ 1 are $\chi^2$ fits
to the corresponding data, based on the functional form
\be
\sigma_x^{\i N}(s) = A_x \left\{ 1 - \left( {s_0^x \over s} \right)^{1/2}
\right\}^{k_x},
\label{8}
\ee
where $x$ refers to elastic and inelastic, respectively, and $s_0^x$
denotes the corresponding threshold energy in each case. The parameters
obtained are given in Table 1.

\begin{figure}[p]
\centerline{\epsfig{file=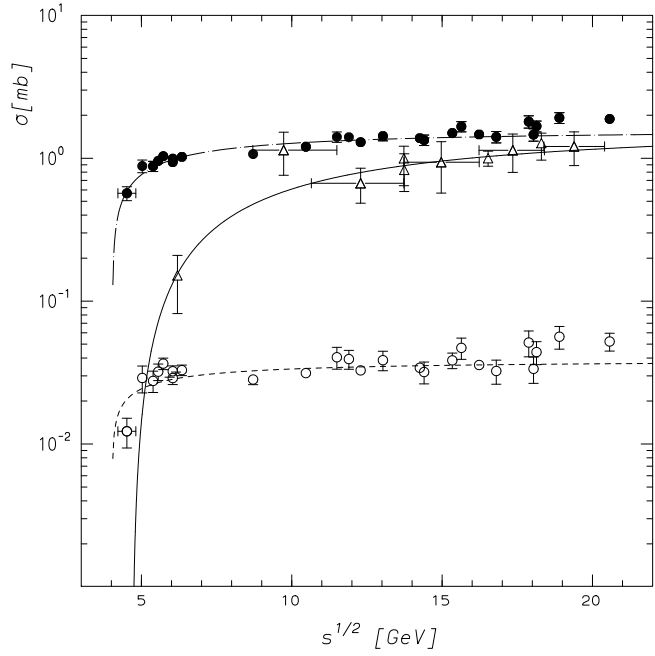,width=90mm}}
\vspace*{-2.0 cm}
\caption{Cross sections for \J-nucleon interactions as obtained
from $J/\psi$ and open charm photoproduction:
$ \sigma_{\rm el}^{\i N}(s)$ (open circles),
$ \sigma_{\rm in}^{\i N}(s)$ (triangles), and
$(1 + \rho^2)^{1/2} \sigma_{\rm tot}^{\i N}(s)$ (filled circles).
The lines give the results of fits (see text).}
\end{figure}

\bigskip

\begin{figure}[p]
\vspace{-1cm}
\centerline{\epsfig{file=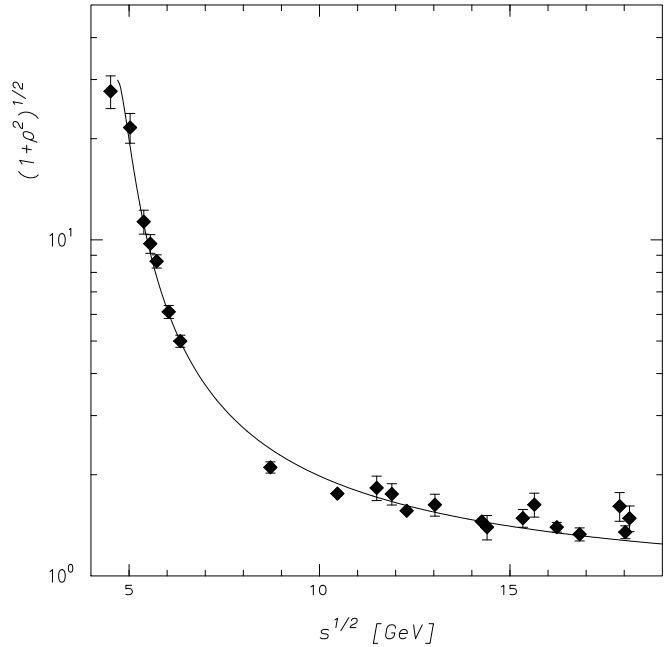,width=90mm}}
\vspace*{-2.0 cm}
\caption{The data and polynomial fit to $(1 + \rho^2)^{1/2}$.}
\end{figure}

Dividing the data for $(1 + \rho^2)^{1/2}\sigma_{\rm tot}^{\i N}(s)$ by
the fitted forms $\sigma_{\rm in}^{\i N}(s) + \sigma_{\rm el}^{\i N}(s)$,
we obtain the energy dependence of the ratio of real to imaginary
parts of the $\i\_N$ scattering amplitude.
This is shown in Fig.\ 2, together with a polynomial fit.
We see that the conditions for the application of
geometric considerations are indeed quite well satisfied for
$\sqrt s~ \gsim~ 15$ GeV, while for $\sqrt s~ \lsim~ 15$ GeV
there are significant deviations. --
Combining the fits of $\sigma_{\rm in}^{\i N}(s)$, $\sigma_{\rm el}
^{\i N}(s)$ and $(1 + \rho^2)^{1/2}$, we obtain a fit to
$(1 + \rho^2)^{1/2}\sigma_{\rm tot}^{\i N}(s)$
(included in Fig.\ 1) which is compatible with the form of
Eq.\ (\ref{8}) and the parameters given in Table 1.

\medskip

\begin{center}
\begin{tabular}{|c|c|c|c|}\hline
$\sigma_x$ & $A_x$ & $k_x$ & ${\chi^2}/{\rm d.o.f.}$ \\ \hline\hline
$\sigma_{\rm in}$ & $1.90 \pm 0.35$ & $1.93 \pm 0.4$ & 0.29 \\ \hline
$\sigma_{\rm el}$ & $0.039 \pm 0.0014$ & $0.284 \pm 0.051$ & 1.7 \\ \hline
$\sqrt{1 + \rho^2}~\sigma_{\rm tot}$ & $1.90 \pm 0.35$ & $0.66 \pm 0.03$
& 3.0 \\ \hline
\end{tabular}

\bigskip

\centerline{Table 1: Fit parameters for $\j \_N$  cross
sections}

\end{center}

The quantity of particular interest for \J~suppression in nuclear
collisions is $\sigma_{\rm in}^{\i N}(s)$; its energy dependence as
obtained from photoproduction is shown in more detail in Fig.\ 3.
Since we have not discussed the threshold behaviour of light quark
contributions to Eq.\ (\ref{6}), the curve of Fig.\ 3 represents in
principle only an upper bound. However, $p \_ p$ data as well as
perturbative studies show a strong threshold suppression also for
open charm hadroproduction \cite{HPC}, so that $\sigma_{\rm in}^{\i
N}(s)$ may well coincide with this upper bound.

\begin{figure}[p]
\begin{center}
\epsfig{file=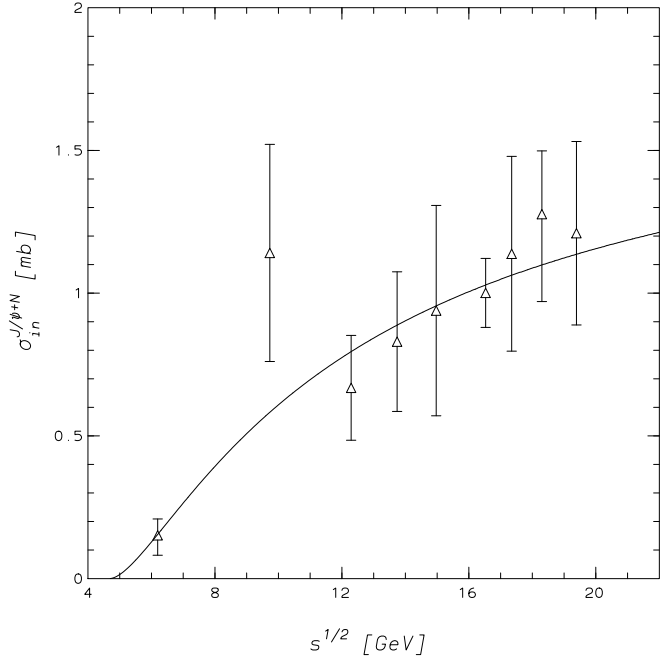,width=90mm}
\end{center}
\vspace*{-2.0 cm}
\caption{The inelastic $J/\psi-N$ cross section together with the fit
given by Eq.\ (\ref{8}).}
\end{figure}

\medskip

\begin{figure}[p]
\begin{center}
\epsfig{file=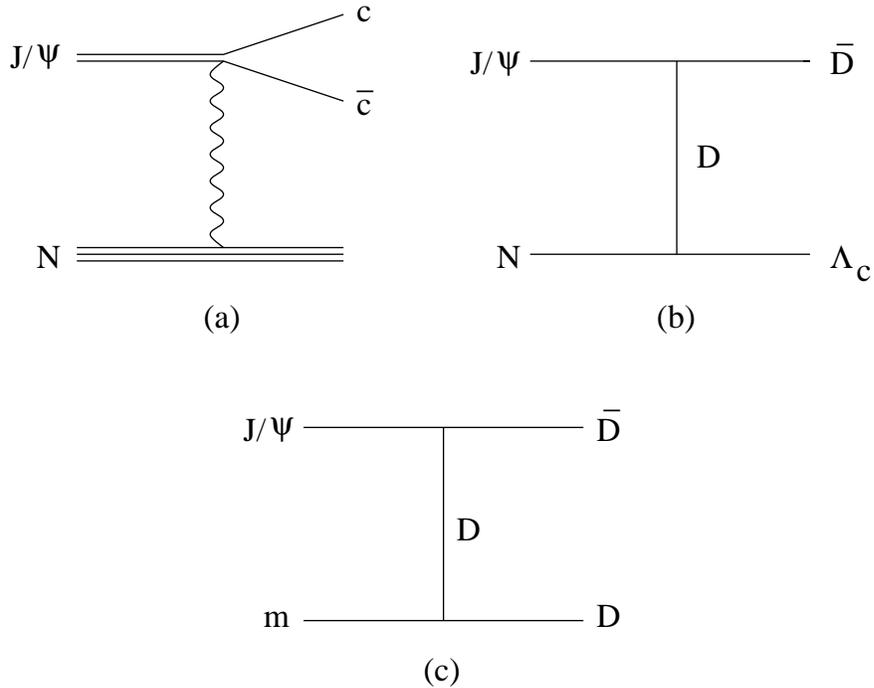, width=90mm, angle=-90}
\end{center}
\caption{Schematic illustrations of 
\J~dissociation by nucleon collisions for (a) short distance QCD
(b) hadron exchange, and (c) hadron exchange in meson collisions.} 
\end{figure}

\medskip

Our considerations are based on vector meson dominance, which assumes
that in \J~photoproduction, a $\C$ fluctuation of a photon of momentum 
$P$ is brought on-shell by interaction with the nucleon, forming a 
\J~of momentum $Q$. For the validity of such a
picture, the longitudinal coherence length $z_L$ of the fluctuation
cannot be much smaller than the size $r_N$ of the nucleon. Hence for
\be
z_L \simeq {1 \over P_L - Q_L} = {1 \over P_L - \sqrt{P_L^2 - M^2_{\i}}}
<<~r_N, \label{10}
\ee
vector meson dominance could break down; we should therefore limit
our results to $\sqrt s ~\gsim~ 5$ GeV in the following discussion.
Note that essentially the entire range shown in Fig.\ 3 falls into the
region of VDM validity.

\medskip

Any model for \J-hadron interactions, whether based on short distance
QCD or on hadron exchange, must satisfy the bound given in Fig's.\ 1
and 3. With this in mind, we now turn to the theoretical
approaches to inelastic $\i\-N$ interactions mentioned above.

\medskip

${\bullet}$
Short distance QCD: The heavy quark constituents and the large binding
energy of the \J~had stimulated short distance QCD calculations quite
some time ago \cite{B-P,SVZ}; these were subsequently elaborated
\cite{Kaidalov} - \cite{KSSZ1}. They are based on the gluon-dissociation
of the \J~(the QCD photo-effect), convoluted with the gluon distribution
function in the nucleon as determined in deep inelastic scattering (see
Fig.\ 4a). The produced final state contains a $D{\bar D}$ pair and a
nucleon, and the resulting form is
\be
\sigma_{\rm in}^{\i N}(s) \simeq \sigma_{\rm in}^{\i
N}(\infty) \left\{ {(2M_D+m)^2 - M_{\i}^2 - m^2 \over s - M_{\i}^2 -
m^2} \right\}^{6.5} \label{11}
\ee
where $\sigma_{\rm in}^{\i N}(\infty)$ denotes the high enery geometric
cross section and $m$ the nucleon mass. Eq.\ (\ref{11}) shows a very
strong damping in the threshold region. The power 6.5 of the damping
factor is obtained from scaling gluon distribution functions; more
realistic distributions will lead to 
a further damping at low and an increase at high $\sqrt s$ \cite{KSSZ2}.

\medskip

${\bullet}$
Charm exchange: The interaction of a \J~with a meson or nucleon
is here considered to take place through open charm exchange.
Such a mechanism has been considered in \cite{M&M} - \cite{Ko} for
\J-meson and \J-nucleon interactions; for the latter it leads to a
  $\Lambda_c$ and a $\bar D$ (see Fig.\ 4b), for the former to a
$D {\bar D}$ final state (Fig. 4c). In the threshold region, the cross
sections for meson ($m$) and nucleon ($N$) projectiles are of
comparable size, as expected from the fact that the ratio of the
couplings
\be
g^2_{DN\Lambda_c}/g^2_{mD{\bar D}} \label{12}
\ee
is of order unity \cite{Haglin}. In \cite{M&M,Ko}, no explicit
results are given for the \J-nucleon cross section. The values obtained
there for \J-meson interactions are quite similar, however, to those in
\cite{Haglin}, where the \J-nucleon interaction is calculated as well.
We shall therefore use this form for our actual comparison.

\medskip

The short distance QCD form Eq.\ (\ref{11}) for inelastic \J-nucleon
interactions, with $\sigma_{\rm in}^{\i N}(\infty)=1.9$ mb,
is seen in Fig.\ 5 to agree quite well with the constraint
from open charm photoproduction. We recall moreover that the 
use of more realistic parton distribution functions would further
improve the agreement. In contrast, the charm exchange cross
section \cite{Haglin} is found to overshoot the
data by more than a factor two over the entire threshold region; the
data point at $\sqrt s=6$ GeV is an order of magnitude lower than the
predicted value. Moreover, the predicted functional form differs from
that of the data. The form shown in Fig.\ 5 is obtained by
smoothly extrapolating the results given in \cite{Haglin} for $\sqrt s
\leq 6$ GeV to the same geometric cross section $\sigma_{\rm
in}^{\i N}(\infty)$ as for the short distance QCD result.

\vspace{-1.5cm}
\begin{figure}[htb]
\begin{center}
\epsfig{file=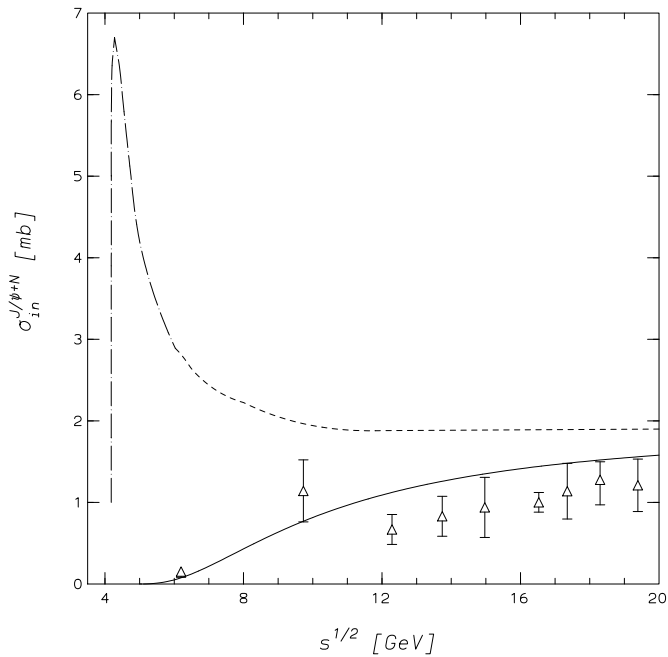, width=90mm}
\end{center}
\vspace*{-2.0 cm}
\caption{The inelastic $J/\psi +N$ cross section compared to
the predictions of the short distance QCD \cite{KS3}
(full line) and the meson exchange model \cite{Haglin} (dashed-dotted line)
extrapolated to higher energy (dashed line).}
\end{figure}

\bigskip

We therefore conclude that the threshold enhancement obtained in hadron
exchange models for inelastic \J-hadron interactions is not compatible
with \J~and open charm photoproduction data. This excludes such
mechanisms as possible source for any `anomalous' \J~suppression
observed in $Pb-Pb$ collisions at the CERN-SPS \cite{NA50}.
Nevertheless, it would be interesting to compare the inelastic
\J-nucleon cross section obtained from photoproduction to possible
direct measurements using either an inverse kinematics \cite{KS5} or
an ${\bar p}A$ annihilation \cite{Dima} experiment.

\medskip

In closing, we note that in addition to these models considered here,
quark interchange or rearrangement has been discussed as possible
mechanism for inelastic \J-hadron interactions \cite{Blaschke,Wong}.
This leads to cross sections which are still much larger very close to
threshold; this is a kinematic region in which VDM is not really
reliable. Nevertheless, the extremely large dissociation cross section
of these models corresponds to a large imaginary part of the \J-hadron
scattering amplitude. Dispersion relations relate its value near
threshold to the real part of the amplitude over a large range of
energies. This is expected to result in an elastic cross section which
strongly violates the bounds shown in Fig.\ 1, so that also here
photoproduction results will very likely prove to be incompatible also
to such an approach \cite{RSSZ}.

\bigskip

\centerline{\bf Acknowledgements}

\medskip

It is a pleasure to thank D.\ Schildknecht and G.\ Schuler for helpful
comments and discussions.  One of us (K.R) also acknowledges the
partial support of the Gesellschaft f\"ur Schwerionenforschung (GSI)
and  of the Committee for Research Development (KBN).

\bigskip


\end{document}